\documentclass[10pt, journal, final, twocolumns]{IEEEtran}
\ifCLASSINFOpdf
\else
\fi
\ifCLASSOPTIONcompsoc
   \usepackage{cite}
\else
   \usepackage{cite}
\fi
 \usepackage[pdftex]{graphicx}
\usepackage{amsmath}
\usepackage{multicol}
\usepackage{graphicx}
\usepackage{epstopdf}
\usepackage{psfrag}
\usepackage{lscape}
\usepackage{amssymb}
\usepackage{amsmath}
\usepackage{cases}
\usepackage{booktabs,multirow}
\usepackage{hyperref}
\usepackage{tikz}
\usetikzlibrary{fit,shapes.misc}

\ifCLASSOPTIONcompsoc
\usepackage[tight,normalsize,sf,SF]{subfigure}
\else
\usepackage[tight,footnotesize]{subfigure}
\fi

\begin{document}
\title{Achieving Economic Operation and Secondary Frequency Regulation Simultaneously Through Feedback Control}

%Interpretations in Dynamics of Dual Decomposition and Method of Multipliers for Power System Economic Operation and Frequency Regulation}

\author{Zhixin Miao,~\IEEEmembership{Senior Member,~IEEE,}
Lingling Fan,~\IEEEmembership{Senior Member,~IEEE}
\thanks{Z. Miao and L. Fan are with Department of Electrical Engineering at University of South Florida, Tampa, FL 33620 (Emails: zmiao@usf.edu, linglingfan@usf.edu)}.}
\maketitle
\begin{abstract}
This article presents an exciting finding for the power industry: the parameters of secondary frequency control based on integral or proportional integral control can be tuned to achieve economic operation and frequency regulation simultaneously.  We show that if the power imbalance is represented by frequency deviation, an iterative dual decomposition based economic dispatch solving is equivalent to integral control. An iterative method of multipliers based economic dispatch is equivalent to proportional integral control. Similarly, if the controller parameters of the secondary frequency controls are chosen based on generator cost functions, these secondary frequency controllers achieve both economic operation and frequency regulation simultaneously.
\end{abstract}

%\maketitle
%\IEEEpeerreviewmaketitle

\section{Introduction}
We will start with a two-area system to explain our finding. In Section II, dual decomposition based iterative economic dispatch problem is presented. The corresponding continuous dynamic model is then derived. In Section III, method of multipliers based iterative approach is presented. The continuous dynamic model is again derived. We show that if the power imbalance can be represented by frequency deviation, the former approach is similar as integral control and the latter approach is similar as proportional integral control.

This finding has a \emph{revolutionary} meaning for the power industry. The parameters of the feedback control for each generator (input: frequency deviation, output: turbine-governor's power reference) can be selected according to generator cost functions. This selection will lead to frequency regulation and economic operation \emph{simultaneously}.

\section{Dual Decomposition Based Iterative Economic Dispatch}
For a two-area system, each area with a generator (dispatch level $P_i$)  and a load ($D_i$), the economic dispatch problem is expressed as follows.
\begin{align}
\min_{P_1, P_2} \>\>\>& C_1(P_1) + C_2(P_2) \notag\\
\text{subject to:} &\>\> P_1+P_2=D_1+D_2.
\end{align}
where $C_i(P_i) = a_iP^2_i+b_iP_i+C_i$ is the cost function.

The dual problem is as follows.
\begin{align}
\max_{\lambda}\>\>\>\>& \min_{P_1} & C_1(P_1) + \lambda(D_1-P_1) \notag\\
  + &   \min_{P_2} \>\>& C_2(P_2) + \lambda(D_2-P_2) 
\end{align}
For a given $\lambda$, Area 1 and Area 2 can carry out minimization problems separately. The price $\lambda$ is then updated to maximize the objective function of the dual problem. In addition, the power imbalance will be represented by the frequency deviation with a gain.
The iterative procedure of $\lambda$ update is as follows.
\begin{align}
\lambda^{k+1} &= \lambda^{k} + \alpha(D_1+D_2-P_1-P_2) \\
              &= \lambda^{k} - K\Delta f^k
\end{align}
Ignoring the generator limits, the marginal costs of the generators should equal to the price at each step.
\begin{align}
\lambda^k = 2a_1P^k_1 + b_1 = 2a_2P^k_2 + b_2.
\end{align}

Therefore, the iteration for the power commands that will be sent to the turbine governors are:
\begin{align}
P^{k+1}_1&=P^k_1 - \frac{K}{2a_1}\Delta f_1^k\\
P^{k+1}_2&=P^k_2 - \frac{K}{2a_2}\Delta f_2^k
\end{align}

The continuous dynamic model of the above procedure can be obtained using Forward Euler approximation for derivatives.
\begin{align}
\tau \dot{P_1} &= - \frac{K}{2a_1}\Delta f_1\\
\tau \dot{P_2} &= - \frac{K}{2a_2}\Delta f_2
\end{align}
where $\tau$ is the step size of the discrete iteration.

\emph{\textbf{Remarks}}: the continuous dynamic model not only indicates that dual decomposition-based economic dispatch is equivalent to an integrator in secondary frequency control, but also indicates that if the gains of the integral controllers for generators are chosen based on generators' cost functions, the local feedback control can realize economic dispatch and frequency regulation simultaneously.

\section{Method of Multipliers Based Iterative Economic Dispatch}
In method of multipliers, an additional term related to an equality constraint is added to the objective function. The advantage of method of multipliers is to achieve faster convergence compared to dual accent method \cite{boyd2011distributed}. Our economic dispatch problem now becomes:
\begin{align}
\min \>\>\>& C_1(P_1)+C_2(P_2) +\frac{\rho}{2}(P_1+P_2-D_1-D_2)^2\\
\text{subject to:}\>& P_1+P_2=D_1+D_2.
\end{align}

Again, the power imbalance can be reflected by frequency deviation. The $\lambda$ update procedure now becomes
\begin{align}
\lambda^{k+1} &= \lambda^k +\rho (D_1+D_2-P^k_1-P^k_2)\notag\\
& = \lambda^k -K\Delta f.
\label{eq:lambda}
\end{align}

For a given $\lambda^k$, $P^k_1$ and $P^k_2$ should minimize the following objective function:
\begin{align}
L(P_1, P_2)  &= C_1(P_1)+C_2(P_2) +  +\frac{\rho}{2}(P_1+P_2-D_1-D_2)^2 \notag \\
&+ \lambda^k(D_1+D_2-P_1-P_2).
\end{align}

The arguments that minimize $L(P_1, P_2)$ can be found by setting the gradients related to $P_1$ and $P_2$ to zeros.
\begin{align}
\frac{\partial L}{\partial P_1}& =0 \notag\\
\frac{\partial L}{\partial P_2}& =0
\end{align}

For the two-area system, we find the relationship of $P^k_1$, $P^k_2$ versus $\lambda^k$ as follows.
\begin{align}
\lambda^k &= 2a_1P^k_1 + b_1 - \rho(D_1+D_2-P^k_1-P^k_2)\\
&= 2a_2P^k_2 + b_2 - \rho(D_1+D_2-P^k_1-P^k_2)\\
\lambda^k & = 2a_1P^k_1 + b_1 +K\Delta f_1 \\
& = 2a_2P^k_2 + b_2 +K\Delta f_2
\label{eq:P1P2}
\end{align}

The iteration for the power commands can be found by replacing $\lambda$ in \eqref{eq:lambda} by \eqref{eq:P1P2}.
\begin{align}
2a_1(P^{k+1}_1-P^{k}_1) +K(\Delta f^{k+1} -\Delta f^k) &= -K\Delta f^k \notag\\
2a_2(P^{k+1}_2-P^{k}_2) +K(\Delta f^{k+1} -\Delta f^k) &= -K\Delta f^k.
\end{align}

Using the Forward Euler approximation for derivatives, the continuous dynamic model is the found to be:
\begin{align}
2a_1 \tau \dot{P_1} + K\tau \dot{\Delta f} =  -K\Delta f_1 \notag\\
2a_2 \tau \dot{P_2} + K\tau \dot{\Delta f} =  -K\Delta f_2.
\end{align}

The transfer functions for the above model are as follows.
\begin{align}
\frac{P_1}{\Delta f_1}&  = \frac{-K}{2a_1}\left(1+\frac{1}{\tau s}\right) \\
\frac{P_2}{\Delta f_2}&  = \frac{-K}{2a_2}\left(1+\frac{1}{\tau s}\right)
\end{align}

\emph{\textbf{Remarks}}: the continuous dynamic model not only indicates that method of multiplers-based economic dispatch is equivalent to a proportional integral (PI) controller in secondary frequency control, but also indicates that if the gains of the PI controllers for generators are chosen based on generators' cost functions, the local feedback control can realize economic dispatch and frequency regulation simultaneously.

The two dynamic models also indicate that method of multiplier-based iteration can achieve faster convergence compared to dual decomposition-based iteration. The former is viewed as a PI control while the latter is viewed as an integral controller. The PI controller should lead to faster response than the integral controller.

%\section{Numerical Results}
%\subsection{Discrete decision making}

%\subsection{Integral and Proportional Integral control for secondary frequency control}

\section{Conclusion}
In this article, we claimed findings of important practice value for the power industry. % explain why method of multiplier based iteration can converge much faster compared to the dual decomposition based iteration from the point of view of dynamic control.
The finding is regarding secondary frequency control: the parameters of the controls can be set based on generators' cost functions to realize economic operation.
We discovered the findings by examining the approximate continuous models for two iterative approaches: method of multipliers and dual decomposition. The former is equivalent to PI control while the latter is equivalent to integral control.

\bibliographystyle{IEEEtran}
\bibliography{IEEEabrv,mas_hybrid}

% Generated by IEEEtran.bst, version: 1.13 (2008/09/30)
\begin{thebibliography}{1}
\providecommand{\url}[1]{#1}
\csname url@samestyle\endcsname
\providecommand{\newblock}{\relax}
\providecommand{\bibinfo}[2]{#2}
\providecommand{\BIBentrySTDinterwordspacing}{\spaceskip=0pt\relax}
\providecommand{\BIBentryALTinterwordstretchfactor}{4}
\providecommand{\BIBentryALTinterwordspacing}{\spaceskip=\fontdimen2\font plus
\BIBentryALTinterwordstretchfactor\fontdimen3\font minus
  \fontdimen4\font\relax}
\providecommand{\BIBforeignlanguage}[2]{{%
\expandafter\ifx\csname l@#1\endcsname\relax
\typeout{** WARNING: IEEEtran.bst: No hyphenation pattern has been}%
\typeout{** loaded for the language `#1'. Using the pattern for}%
\typeout{** the default language instead.}%
\else
\language=\csname l@#1\endcsname
\fi
#2}}
\providecommand{\BIBdecl}{\relax}
\BIBdecl

\bibitem{boyd2011distributed}
S.~Boyd, N.~Parikh, E.~Chu, B.~Peleato, and J.~Eckstein, ``Distributed
  optimization and statistical learning via the alternating direction method of
  multipliers,'' \emph{Foundations and Trends{\textregistered} in Machine
  Learning}, vol.~3, no.~1, pp. 1--122, 2011.

\end{thebibliography}
%\vspace{-0.3in}
%\begin{biographynophoto}{Javad Khazaei}
%is a Ph.D student at University of South Florida (USF). He received his Bachelor degree in Electrical Engineering from Mazandaran University (2009) and Master degree from Urmia University (2011) in Iran. He started his Ph.D study at USF in Summer 2013 and his research interests include Smart Grid modeling, Renewable Energy Integration, and Power Electronics.
%\end{biographynophoto}
%
\vspace{-0.3in}
%\begin{IEEEbiography} [{\includegraphics[width=1in,height=1.25in]{zmiao}}]{Zhixin Miao}(S'00 M'03 SM'09) received the
\begin{IEEEbiographynophoto}{Zhixin Miao}(S'00 M'03 SM'09) received the
B.S.E.E. degree from the Huazhong University of
Science and Technology, Wuhan, China, in 1992, the
M.S.E.E. degree from the Graduate School, Nanjing
Automation Research Institute, Nanjing, China, in
1997, and the Ph.D. degree in electrical engineering
from West Virginia University, Morgantown, in
2002.

Currently, he is with the University of South
Florida (USF), Tampa. Prior to joining USF in 2009,
he was with the Transmission Asset Management
Department with Midwest ISO, St. Paul, MN, from 2002 to 2009. His research
interests include power system stability, microgrid, and renewable energy.
%\end{IEEEbiography}
\end{IEEEbiographynophoto}
\begin{IEEEbiographynophoto}{Lingling Fan}
%\begin{IEEEbiography}[{\includegraphics[width=1in,height=1.25in]{lfan3}}]{Lingling Fan}(SM'08)
%%\begin{biographynophoto}{Lingling Fan}
received the B.S. and M.S. degrees in electrical engineering from Southeast University,
Nanjing, China, in 1994 and 1997, respectively, and the Ph.D. degree in electrical engineering
from West Virginia University, Morgantown, in 2001.
Currently, she is an Associate Professor with the University of South Florida, Tampa, where she has
been since 2009. She was a Senior Engineer in the Transmission Asset Management Department, Midwest
ISO, St. Paul, MN, form 2001 to 2007, and an Assistant Professor with North Dakota State University,
Fargo, from 2007 to 2009. Her research interests include power systems and power electronics. Dr. Fan serves as a technical program committee chair for IEEE Power System Dynamic Performance Committee and an editor for IEEE Trans. Sustainable Energy.
%\end{IEEEbiography}
\end{IEEEbiographynophoto}

\end{document}